\documentclass[aps,preprintnumbers,amsmath,amssymb,twocolumn, tightenlines,superscriptaddress]{revtex4}

\usepackage{graphicx}
\usepackage{epsfig}

\newcommand{\be}{\begin{eqnarray}}
\newcommand{\ee}{\end{eqnarray}}

 \newcommand{\gsim}{\mathrel{\hbox{\rlap{\lower.55ex \hbox {$\sim$}}
                   \kern-.3em \raise.4ex \hbox{$>$}}}}
\newcommand{\lsim}{\mathrel{\hbox{\rlap{\lower.55ex \hbox {$\sim$}}
                   \kern-.3em \raise.4ex \hbox{$<$}}}}

\newcommand{\ba}{\begin{eqnarray}}
\newcommand{\ea}{\end{eqnarray}}

\begin{document}


\title{Quantifying Chiral Magnetic Effect from Anomalous-Viscous Fluid Dynamics}
\author{Yin Jiang} 
\address{School of Physics and Nuclear Energy Engineering, Beihang University, Beijing 100191, China.}
\author{Shuzhe Shi} 
\address{Physics Department and Center for Exploration of Energy and Matter,
Indiana University, 2401 N Milo B. Sampson Lane, Bloomington, IN 47408, USA.}
\author{Yi Yin}
\address{Center for Theoretical Physics, Massachusetts Institute of Technology, Cambridge, MA 02139, USA.}
\author {Jinfeng Liao} \email{liaoji@indiana.edu}
\address{Physics Department and Center for Exploration of Energy and Matter,
Indiana University, 2401 N Milo B. Sampson Lane, Bloomington, IN 47408, USA.}
\address{Institute of Particle Physics and Key Laboratory of Quark \& Lepton Physics (MOE), Central China Normal University, Wuhan, 430079, China.}

\begin{abstract}
The Chiral Magnetic Effect (CME) is a macroscopic manifestation of fundamental chiral anomaly in a many-body system of chiral fermions, and emerges as anomalous transport current in the fluid dynamics framework.  Experimental observation of CME is of great interest and has been reported in Dirac and Weyl semimetals. Significant efforts have also been made to look for CME in heavy ion collisions. 
Critically needed for such search, is the theoretical prediction for CME signal. In this paper we report a first quantitative modeling framework, the Anomalous Viscous Fluid Dynamics (AVFD), which computes the evolution of fermion currents on top of realistic bulk evolution in heavy ion collisions and simultaneously accounts for both anomalous and normal viscous transport effects. The AVFD allows a quantitative understanding of the generation and evolution of CME-induced charge separation during  hydrodynamic stage  as well as its dependence on theoretical ingredients. With reasonable estimates of key parameters, the AVFD simulations provide the first phenomenologically successful explanation of the measured signal in 200AGeV AuAu collisions. 
\end{abstract}

\maketitle


{\it Introduction.---} The importance of electricity for modern society cannot be overemphasized. From the physics point of view, lies at the heart of electricity is the conducting transport (of electric charge carriers). In normal materials, conducting transport generates an electric current $\vec{\bf J}_Q$ along the electric field $\vec{\bf E}$ (or voltage)  applied to the system. This can be described by the usual Ohm's law $\vec{\bf J}_Q = \sigma_e  \vec{\bf E}$ where the conductivity $\sigma_e$ arises from competition between ``ordered'' electric force and ``disordered'' thermal scatterings, henceforth involving dissipation and typically dependent upon specific dynamics of the system. More recently there have been significant interests,  from both high energy and condensed matter physics communities, in a new category of  {\it anomalous chiral transport} in quantum materials containing chiral fermions. A  notable example is the   Chiral Magnetic Effect (CME)~\cite{Vilenkin:1980fu,Kharzeev:2004ey,Kharzeev:2007tn,Kharzeev:2007jp,Fukushima:2008xe} --- the generation of an electric current    $\vec{\bf J}_Q$ along the {\it magnetic field} $\vec{\bf B}$ applied to the system, i.e.
\begin{eqnarray} \label{eq_cme}
\vec{\bf J}_Q = \sigma_5 \vec{\bf B}
\end{eqnarray} 
where $\sigma_5 = C_A \mu_5$ is the chiral magnetic conductivity, expressed in terms of the chiral chemical potential $\mu_5$ that quantifies the imbalance between fermions of opposite (right-handed, RH versus left-handed, LH) chirality. 

The $\sigma_5$ has two remarkable features that make it markedly different from the normal conductivity $\sigma_e$. First, the coefficient 
$C_A$ takes a {\it universal value} of $Q_f^2/(4\pi^2)$ (for each species of RH or LH fermions with electric charge $Q_f$) from non-interacting cases to extremely strongly coupled cases~ \cite{Fukushima:2008xe,Son:2009tf,Zakharov:2012vv,Fukushima:2012vr}. In fact, it is entirely dictated by universal chiral anomaly coefficient,  and the CME is really just the macroscopic manifestation of the fundamental quantum anomaly in a many-body setting. Second, the $\sigma_5$  is time-reversal even~\cite{Kharzeev:2011ds} which implies the non-dissipative nature of the underlying transport process that leads to the CME current in (\ref{eq_cme}).

Given the magnificent physics of Chiral Magnetic Effect, it is of utmost interest to search for its manifestation in real-world materials. Two types of systems for experimental detection of CME have been enthusiastically investigated. One is the so-called Dirac and Weyl semimetals where electronic states emerge as effective chiral fermions and exhibit chiral anomaly~\cite{Nielsen:1983rb,Son:2012bg}. Discoveries of CME were reported in those systems~\cite{Li:2014bha,Xiong:2015nna,2015PhRvX...5c1023H,2016NatCo...711615A}. The other is the quark-gluon plasma (QGP), which is the deconfined form of nuclear matter at very high temperatures $T\sim$ trillion degrees, consisting of approximately massless light quarks. Such a new form of hot matter once filled the whole universe and is now (re)created in laboratory at the Relativistic Heavy Ion Collider (RHIC) and the Large Hadron Collider (LHC). Search for potential CME signals has been ongoing at RHIC and the LHC~\cite{STAR_LPV1,STAR_LPV2,STAR_LPV3,STAR_LPV4,STAR_LPV_BES,ALICE_LPV}, with encouraging evidences for CME-induced charge separation signal. The interpretation of these data however suffers from backgrounds arising from the complicated environment in a heavy ion collision (see   e.g.~\cite{Kharzeev:2015znc,Liao:2014ava,STAR_LPV_Wang,Bzdak:2012ia}). Currently the most pressing challenge for the search of CME in heavy ion collisions is to clearly separate background contributions from the desired signal. A mandatory and critically needed step, is to develop state-of-the-art modeling tools to compute CME signal in a realistic heavy ion collision environment. In this Letter we present such a tool, the Anomalous Viscous Fluid Dynamics (AVFD) framework, which simulates the evolution of chiral fermion currents in the QGP on top of the  VISHNU bulk hydrodynamic evolution for heavy ion collisions. We demonstrate the features of  this framework and quantify the CME-induced charge separation signal for comparison with available experimental data.

    \begin{figure*}[hbt!]
\begin{center} 
\includegraphics[scale=0.24]{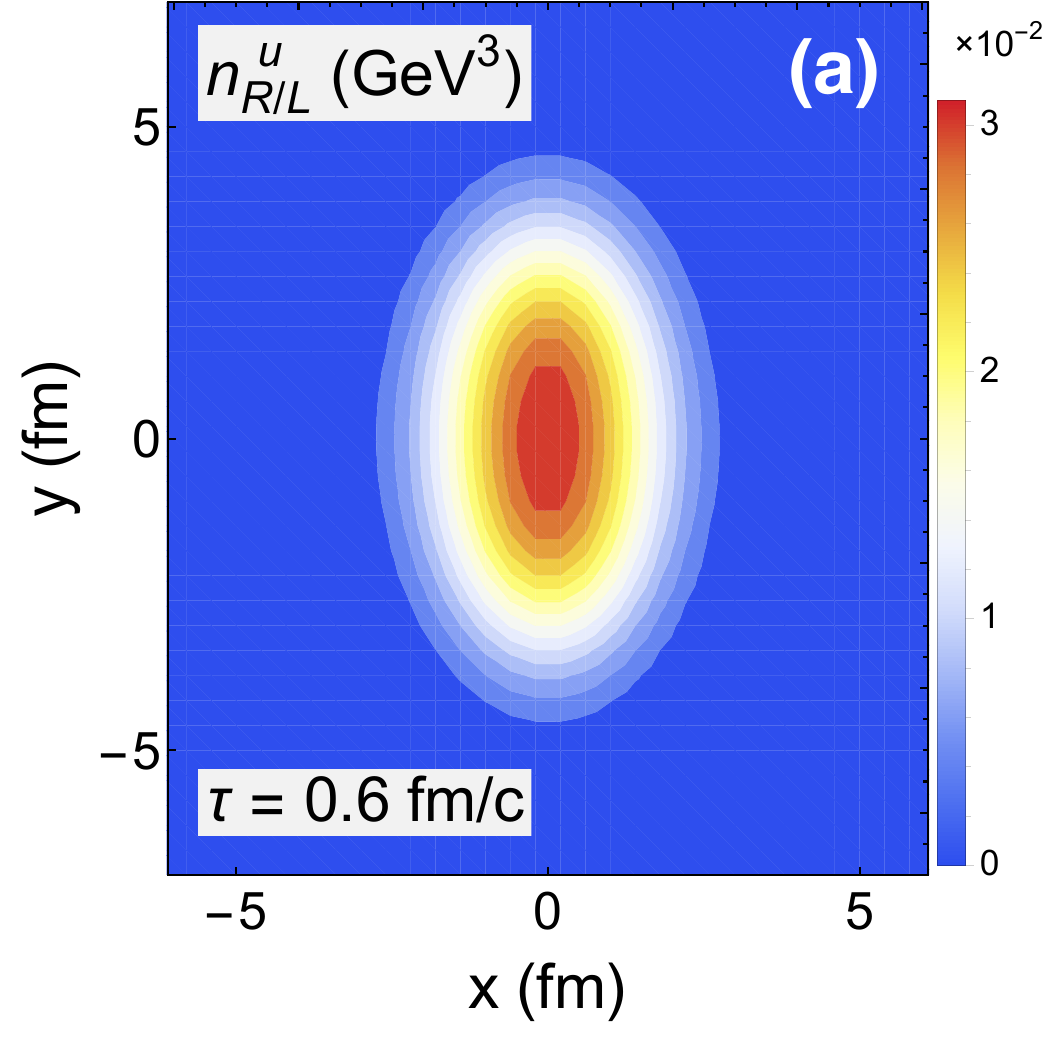} 
\includegraphics[scale=0.24]{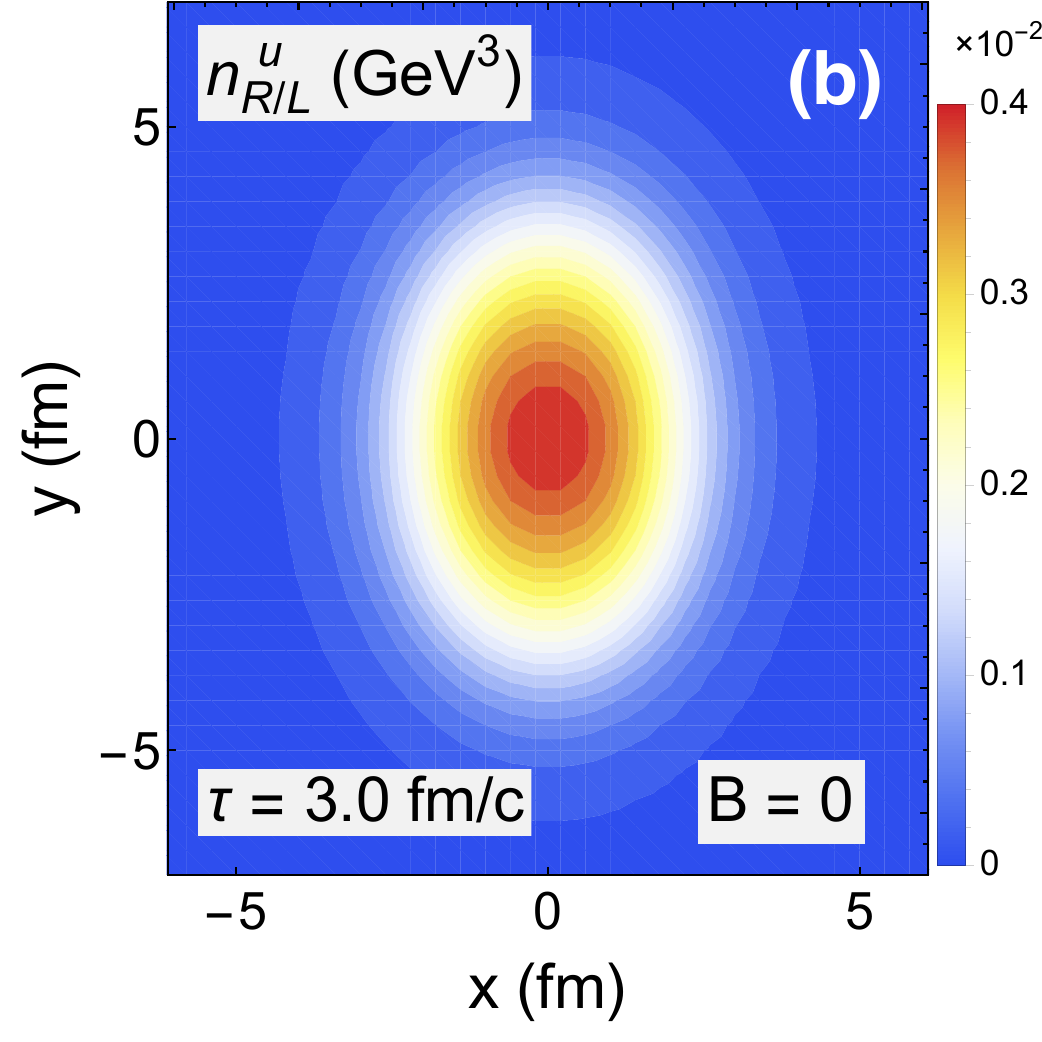}
\includegraphics[scale=0.24]{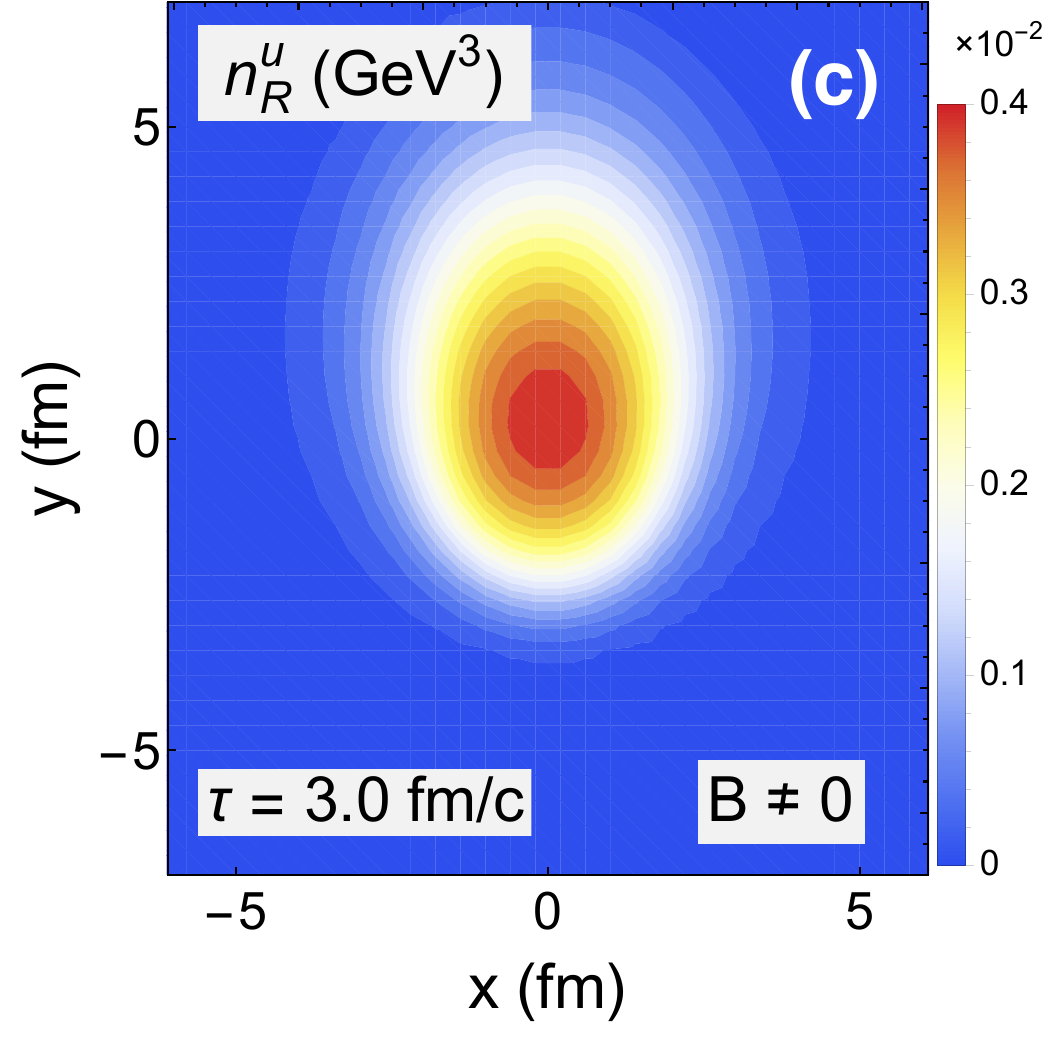} 
\includegraphics[scale=0.24]{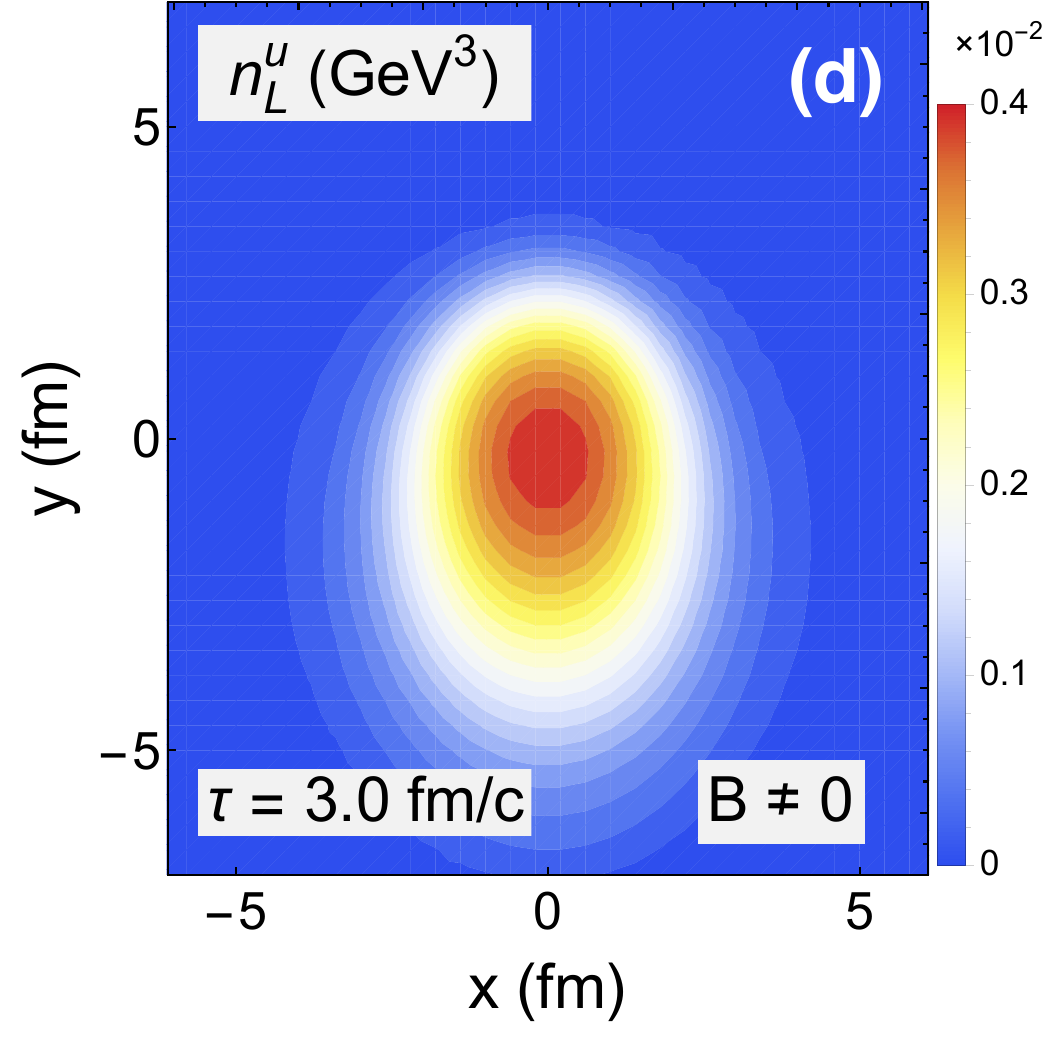}
\caption{(color online) The evolution of u-flavor densities via solving AVFD equations from the same initial charge density distribution (for either RH or LH) at $\tau_0=0.60\rm fm/c$ [panel (a)]   in three different cases: (1) [panel (b)] for either RH or LH density at $\tau=3.00\rm fm/c$ with zero magnetic field ${\bf B}\to 0$ which implies no anomalous chiral transport; (2)  [panel (c)] for RH density at $\tau=3.00\rm fm/c$ with nonzero ${\bf B}$ field along positive y-axis; (3)  [panel (d)] for LH density at $\tau=3.00\rm fm/c$ with nonzero ${\bf B}$ field along positive y-axis. } \label{fig1} \vspace{-0.3in}
\end{center}
\end{figure*}

{\it The Anomalous-Viscous Fluid Dynamics.---}Fluid dynamics provides a universal description of macroscopic systems under the large scale and long time limit, and are essentially conservation laws  arising  from symmetries in microscopic dynamics. For a fluid of chiral fermions, the microscopic chiral anomaly is a sort of ``half symmetry'' and how it arises in macroscopic fluid dynamics is a nontrivial question. As answered in \cite{Son:2009tf}, the constituent relation for the fermion currents is required by the second law of thermodynamics to include anomalous terms corresponding to the CME current and a similar chiral vortical current. Based on such finding, let us then develop a simulation framework to describe anomalous chiral transport in heavy ion collisions at very high beam energy (e.g. top RHIC energy and above). The bulk evolution in such collisions is well described by  2+1D 2nd-order viscous hydrodynamics (e.g. VISHNU simulations~\cite{Shen:2014vra}). We use the ``single-shot'' version with averaged smooth energy-momentum tensor initial condition based on Monte-Carlo Glauber model.   
Our approach is to solve the following fluid dynamical equations for the evolution of chiral fermion currents (RH and LH currents for u and d flavors respectively) as perturbations on top of the bulk fluid evolution: 
\begin{eqnarray} \label{eq_avfd} 
\hat{D}_\mu J_{\chi, f}^\mu &=& \chi \frac{N_c Q_f^2}{4\pi^2} E_\mu B^\mu \\
J_{\chi, f}^\mu &=& n_{\chi, f}\, u^\mu + \nu_{\chi, f}^\mu + \chi \frac{N_c Q_f}{4\pi^2} \mu_{\chi, f} B^\mu 
\label{eq_J}   \\ 
\Delta^{\mu}_{\,\, \nu} \hat{d} \left(\nu_{\chi, f}^\nu \right) &=& - \frac{1}{\tau_{r}} \left[  \left( \nu_{\chi, f}^\mu \right) -  \left(\nu_{\chi, f}^\mu \right)_{NS} \right ] 
\label{eq_ns_1}\\
\left(\nu_{\chi, f}^\mu \right)_{NS} &=&  \frac{\sigma}{2} T \Delta^{\mu\nu}   \partial_\nu \left(\frac{\mu_{\chi, f}}{T}\right) +  \frac{\sigma}{2} Q_f E^\mu   \quad
\label{eq_ns_2}
\end{eqnarray} 
where $\chi=\pm1$ labels chirality for RH/LH currents and $f=u,d$ labels light quark flavor with respective electric charge $Q_f$ and with color factor $N_c=3$. The $E^\mu=F^{\mu\nu}u_\nu$ and  $B^\mu=\frac{1}{2}\epsilon^{\mu\nu\alpha\beta}u_\nu F_{\alpha\beta}$ are   external electromagnetic fields in fluid rest frame.  The derivatives $\hat{D}_\mu$ is covariant derivative and  $\hat{d} =u^\mu \hat{D}_\mu$, with projection operator $\Delta^{\mu \nu}=\left(g^{\mu\nu} - u^\mu u^\nu \right)$.  
In the above the fluid four-velocity field $u^\mu$, the  local temperature $T$ as well as all other thermodynamic quantities are determined by background bulk flow. Furthermore the (small) fermion densities $n_{\chi, f}$ and corresponding chemical potential $ \mu_{\chi, f}$ are related by lattice-computed quark number susceptibilities $c_2^f(T)$~\cite{Borsanyi:2011sw,Bazavov:2012jq}.  Two  transport coefficients are also involved: the normal diffusion coefficient $\sigma$ and the relaxation time $\tau_r$. As the QGP is in a chiral-symmetric phase, we use the same values for susceptibilities and relaxation time for vector and axial charges.  
We set the vector charge density initial condition to be zero and it has been explicitly checked that the anomalous transport effect is insensitive to nonzero initial vector charge density. 
 The above framework treats the normal viscous currents $\nu^\mu_{\chi, f}$ at the 2nd-order of gradient expansion by incorporating relaxation evolution toward Navier-Stocks form. Owing to the quantum nature of the anomalous current, we do not include any 2nd-order thermal relaxation on this contribution.   Note that the axial charge should  ``suffer'' from dissipative effects due to gluonic topological fluctuations and finite quark mass. However the relevant relaxation time scales from both effects have been estimated (see e.g. \cite{Lin:2013sga,Manuel:2015zpa}) to be significantly longer than the $\vec{{\bf B}}$ field lifetime and therefore justify the approximation of neglecting them here. To compute final hadron observables, we convert vector charge densities on the hydro freeze-out surface via susceptibilities into corresponding chemical potentials, which are then incorporated into the thermal distribution of the standard Cooper-Frye procedure.

The most unique feature of the above Anomalous-Viscous Fluid Dynamics (AVFD) framework lies in the ${\bf B}$-field driven anomalous current --- the last term in Eq.(\ref{eq_J}) which distinguishes the left from the right with opposite sign. We demonstrate the effect of such chiral transport in Fig.\ref{fig1}, by computing the evolution of u-flavor currents via solving AVFD equations from the same initial charge density distribution (for either RH or LH) at $\tau_0=0.60\rm fm/c$  [panel (a)]   in three different cases: (1)  [panel (b)] for either RH or LH density at $\tau=3.00\rm fm/c$ with zero magnetic field ${\bf B}\to 0$ which implies no anomalous chiral transport; (2)  [panel (c)] for RH density at $\tau=3.00\rm fm/c$ with nonzero ${\bf B}$ field along positive y-axis; (c)  [panel (d)] for LH density at $\tau=3.00\rm fm/c$ with nonzero ${\bf B}$ field along positive y-axis. In the case (a) the densities evolve only according to normal viscous transport i.e. charge diffusion which is identical  for RH/LH densities and ``up/down'' symmetric. Under the presence of ${\bf B}$ field, additional transport occurs via anomalous currents along the ${\bf B}$ field direction, and RH/LH densities evolve in an asymmetric and opposite way.  The effect of ${\bf B}$-field driven anomalous chiral transport is evident from such a comparison. 

{\it CME-Induced Charge Separation.---} With the AVFD simulation tool introduced above, we are now ready to quantify the CME-induced charge separation signal under realistic conditions in heavy ion collisions. 
Such a charge separation arises from the anomalous current along $\vec{{\bf B}}$ field direction and leads to a dipole term in the azimuthal distribution of produced charged hadrons: 
\begin{eqnarray}
\frac{dN^{ch}}{d\phi} \propto [1\pm 2 a^{ch}_1 \sin\phi   + ...]
\end{eqnarray}
where $\phi$ is the azimuthal angle measured with respect to the reaction plane, and the $\pm a^{ch}_1$ for  opposite charges respectively. 
The charge separation signal critically depends upon the magnetic field and initial axial charge, both of which are not theoretically well constrained. 

For the magnetic field $\vec {{\bf B}} = {\bf B}(\tau) \hat{y}$ (with $\hat{y}$ the event-wise out-of-plane direction), we use a plausible parametrization~\cite{Yin:2015fca,Yee:2013cya}  
\begin{eqnarray}
{\bf B}(\tau) = \frac{{\bf B}_0}{1+\left(\tau / \tau_B\right)^2}
\end{eqnarray}
The peak value ${\bf B}_0$ (for each centrality)  has been well quantified with event-by-event simulations and we use the event-plane projected realistic values from \cite{Bloczynski:2012en} which properly takes into account the misalignment between $\vec{{\bf B}}$ field direction and event plane due to fluctuations. The lifetime $\tau_B$ is poorly known~\cite{McLerran:2013hla,Gursoy:2014aka,Tuchin:2015oka}. Logically there are three possibilities: (a) $\tau_B$ is much longer than the initial  time of hydrodynamic evolution $\tau_0=0.6\rm fm/c$ which appears unlikely; (b) $\tau_B$ is extremely short, $\tau_B\ll \tau_0$, in which case the anomalous chiral transport would have to occur in an out-of-equilibrium setting; (c) $\tau_B$ is short but comparable to $\tau_0$, i.e. $\tau_B\simeq \tau_0$, which is a plausible assumption we adopt for the present work. 

For the initial axial charge density arising from gluonic topological charge fluctuations, one could make an  estimate based on the strong chromo-electromagnetic fields in the early-stage glasma (see e.g. \cite{Kharzeev:2001ev,Muller:2010jd,Hirono:2014oda,Mace:2016svc}): 
\begin{eqnarray} \label{eq_n5}
\sqrt{ \left< n_5^2 \right> } \simeq  \frac{Q_s^4\, (\pi \rho_{tube}^2 \tau_0) \, \sqrt{N_{coll.}}}{16\pi^2 \, A_{overlap}}
\end{eqnarray}
In the above $\rho_{tube}\simeq 1 \rm fm$ is the transverse extension of glasma flux tube, $A_{overlap}$ is the geometric overlapping area of the two colliding nuclei, and $N_{coll.}$ the binary collision number for a given centrality. The above estimate is then used to determine a ratio $\lambda_5$ of total average axial charge over the total entropy in the fireball at initial time $\tau_0$, $\lambda_5 \equiv \frac{\int_V \, \sqrt{ \left< n_5^2 \right> }}{\int_V \, s}$ where the integration is over fireball spatial volume and the $s$ is the entropy density from bulk hydro initial condition. This ratio is then used in the AVFD simulations to set an   initial axial charge distribution locally proportional to entropy density via  $n_5^{initial} = \lambda_5 \, s$. This properly reflects the fact that axial charge arises from local domains with gluon topological fluctuations and that there are more such domains where the matter is denser. 
Such initial axial charge condition depends most sensitively upon the saturation scale $Q_s$, in the reasonable range of  $Q_s^2\simeq 1\sim 1.5 \rm GeV^2$ for RHIC 200AGeV collisions~\cite{Rezaeian:2012ji,Schenke:2012wb,Kowalski:2007rw}. These estimates are consistent with the first  first-principle simulations of off-equilibrium sphelaron transition rates~\cite{Mace:2016svc}.

\begin{figure}[!hbt]
\begin{center}
\includegraphics[scale=0.45]{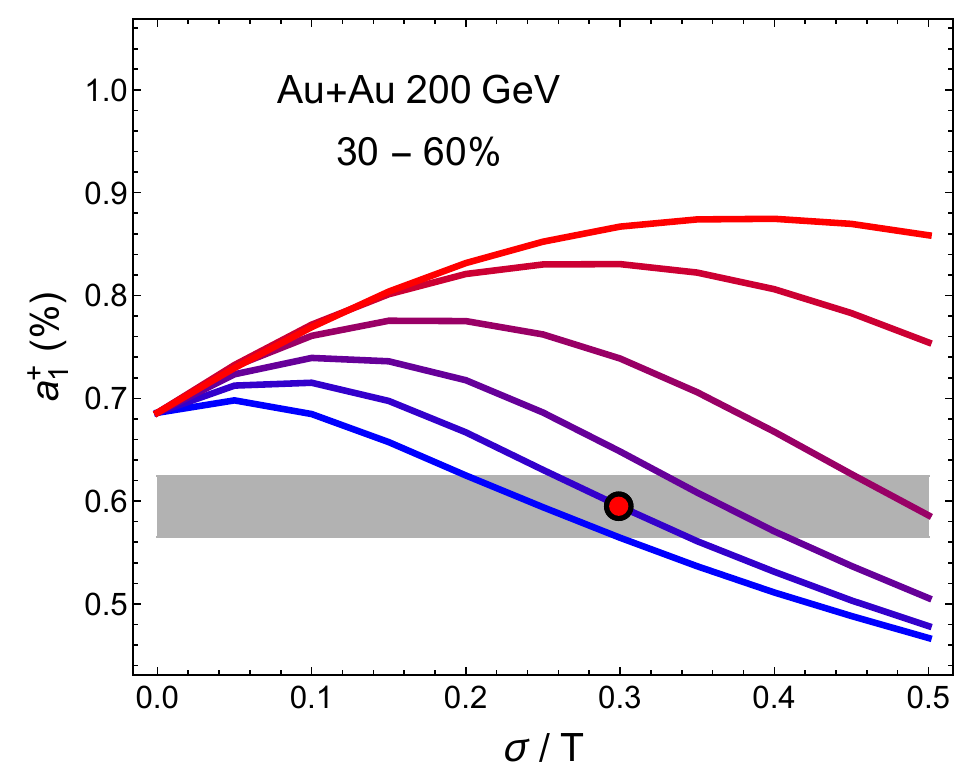}
\vspace{-0.1in}
\caption{(color online) The charge separation signal $a^{ch}_1$ computed from AVFD for $30\sim 60\%$ centrality as a function of diffusion coefficient $\sigma/T$ for given relaxation time $\tau_r T=0.3, 0.5, 0.7, 1.0, 1.5, 2.0$ (from bottom/blue to top/red) respectively. The red blob indicates the result for $\sigma/T=0.3$ and $\tau_r T=0.5$, with the grey shaded band indicating a $\pm 5\%$ deviation in $a^{ch}_1$ from this choice.}
\vspace{-0.15in}
\label{fig2}
\end{center}
\end{figure}

In addition there are two important viscous transport parameters, the diffusion coefficient $\sigma$ and the relaxation time $\tau_r$, the values of which are not precisely determined yet, albeit narrowed down to certain plausible choice for the QGP in the relevant temperature regime. For the quantitative study of CME, 
it is crucial to understand the dependence of the anomalous transport on these normal viscous parameters and to characterize the associated theoretical uncertainty. This has not been possible in  early attempts of anomalous transport modelings in the ideal hydrodynamic limit~\cite{Yin:2015fca,Yee:2013cya,Hirono:2014oda,Hongo:2013cqa}. The AVFD for the first time provides a tool to fully address such question. In Fig.~\ref{fig2} we show the computed charge separation signal $a^{ch}_1$ for one centrality bin ($30\sim 60\%$) versus conductivity $\sigma/T$ at various choices of $\tau_r T$ with $T$ the temperature. Within a relatively wide range of values for $\sigma/T$ and $\tau_r T$, the resulting $a^{ch}_1$ varies from the ideal-hydro-limit (for $\sigma\to 0$) within about $\pm 30\%$ range.  A ``canonical choice'' (see e.g.~\cite{Denicol:2012vq}) of $\sigma/T=0.3$ and $\tau_r T=0.5$, which will be used in our later computation, is indicated by the red blob, with the grey shaded band indicating a $\pm 5\%$ deviation in $a^{ch}_1$ from this choice.

\begin{figure}[!htbt]
\begin{center}
\includegraphics[scale=0.4]{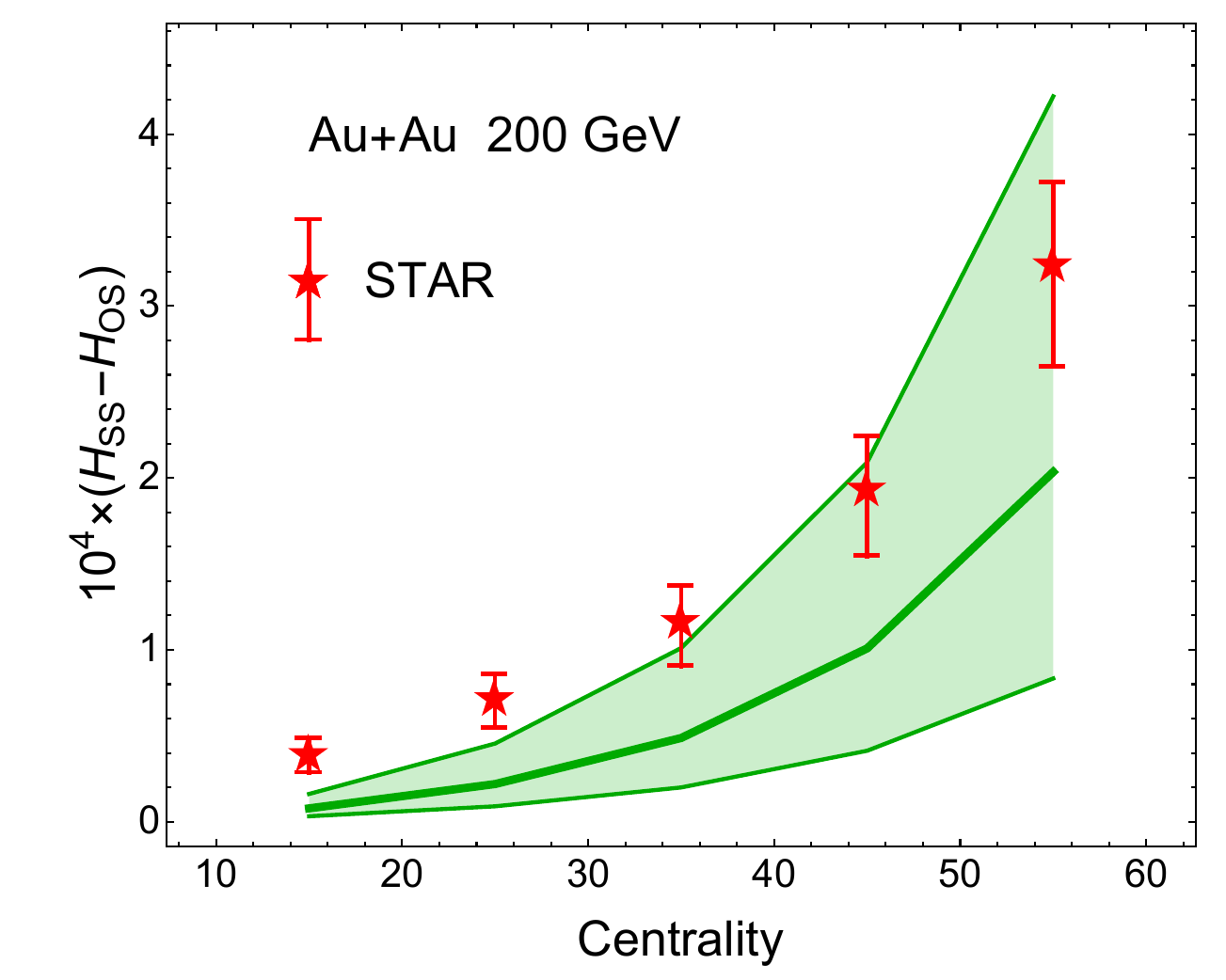} 
\vspace{-0.1in}
\caption{(color online) The azimuthal correlation observable $\left(H_{SS}-H_{OS}\right) $ for various centrality, computed from AVFD simulations and compared with STAR data~\cite{STAR_LPV_BES},  with the green band spanning the range of key parameter from $Q_s^2=1\rm GeV^2$ (bottom edge) to  $Q_s^2= 1.5 \rm GeV^2$ (top edge). }
\vspace{-0.3in}
\label{fig3}
\end{center}
\end{figure}

Let us now utilize the AVFD tool for quantifying CME toward comparison with available data. The measurement of a CME-induced charge separation is however tricky, as this dipole flips its sign  from event to event pending the sign of the initial axial charge arising from fluctuations, thus with a vanishing event-averaged mean value. What can be measured is its variance, through azimuthal correlations for same-sign (SS) and opposite-sign (OS) pairs of charged hadrons~\cite{Voloshin:2004vk,STAR_LPV1,STAR_LPV2}. The so-called $\gamma_{SS/OS}\equiv \left< \cos(\phi_1+\phi_2)\right>$ observables measure a difference between the in-plane versus out-of-plane correlations and are indeed sensitive to potential CME contributions. They however suffer from considerable flow-driven background contributions that are not related to CME: see \cite{Kharzeev:2015znc,Bzdak:2012ia} for reviews and references. A lot of efforts have been made, attempting to separate backgrounds from CME signals (see most recent discussions in e.g. \cite{Bloczynski:2013mca,Skokov:2016yrj,Deng:2016knn,Wang:2016iov,Wen:2016zic}). One approach based on a two-component scenario~\cite{Bzdak:2012ia,Bloczynski:2013mca} was recently adopted by the STAR Collaboration to suppress backgrounds and   extract the flow-independent part (referred to as $H_{SS/OS}$)~\cite{STAR_LPV_BES}. We consider $H_{SS/OS}$ as our ``best guess'' thus far for potential CME signal to be compared with AVFD computations. Specifically a pure CME-induced charge separation will contribute as $\left(H_{SS}-H_{OS}\right) \to 2\left(a^{ch}_1\right)^2$. The AVFD results for various centrality bins are presented in Fig.~\ref{fig3}, with the green band spanning the range of key parameter $Q_s^2$ in $1\sim 1.5 \rm GeV^2$  reflecting uncertainty in estimating initial axial charge (see Eq.(\ref{eq_n5})). Clearly the CME-induced correlation is very sensitive to the amount of initial axial charge density as controlled by $Q_s^2$. The comparison with STAR data~\cite{STAR_LPV_BES} shows quantitative agreement for the magnitude and centrality trend for choices with relatively large values of $Q_s^2$. Therefore the AVFD simulations provide the first phenomenologically successful explanation of the measured signal. That said, it is useful to keep in mind the current uncertainties both in theory (mainly on {\bf B}-field lifetime and initial axial charge) and in experiment (mainly potential residue backgrounds in the $H$-correlation~\cite{Wang:2016iov,Wen:2016zic}). With the AVFD as a versatile tool for quantifying the CME, a definitive conclusion could be expected with future efforts in narrowing down  the theoretical as well as experimental uncertainties.

 If there is considerable CME transport occurring before the start of hydrodynamics, then such pre-hydro CME contribution can be incorporated into the AVFD framework as nontrivial initial conditions for the currents $J^\mu_{\chi, f}$ and  such  pre-hydro charge separation survives into final hadron observables with certain reduction factor, as demonstrated by previous transport study~\cite{Ma:2011uma} and also quantitatively seen in our AVFD simulations.

{\it Summary and Discussions.---} In summary, the CME is a new type of macroscopic anomalous transport arising from microscopic anomaly in chiral matter. Given its observation in Dirac and Weyl semimetal systems in condensed matters experiments, it is now of extreme relevance and significance to  search for such effect  in an entirely different system i.e. the QGP, for the CME as a universal emergent phenomenon. A critically needed theoretical tool for this effort, the Anomalous Viscous Fluid Dynamics (AVFD), has been developed in this paper, which  simultaneously accounts for both anomalous and normal viscous transport effects of fermion currents on top of realistic bulk evolution in heavy ion collisions. The AVFD allows a quantitative understanding of the generation and evolution of CME-induced charge separation during  hydrodynamic stage  as well as its dependence on theoretical ingredients. With reasonable estimates of key parameters, the AVFD simulations have provided the first phenomenologically successful explanation of the measured signal in 200AGeV AuAu collisions.

We end by briefly mentioning a number of interesting problems that are being explored with this new tool. 
These include studying  possible anomalous transport of  fermions with finite mass~\cite{Mueller:2016ven} for quantifying possible charged kaon separation signal, the influence of different magnetic field time dependence, the CME-induced gapless excitation known as the Chiral Magnetic Wave which leads to a splitting in the positive/negative pion elliptic flow~\cite{Burnier:2011bf,Adamczyk:2015eqo}, as well as the development of event-by-event simulations.  These, as well as many other new studies enabled by the AVFD, will be reported in a future publication. 
 
 \vspace{0.1in}

 {\bf Acknowledgments.} 
The authors thank B. Chen, H. Huang, X.-G. Huang, D. Kharzeev, V. Koch, R. Lacey, S. Schlichting, M. Stephanov, A. Tang, R. Venugopalan, F. Wang, G. Wang, N. Xu, H.-U. Yee and P. Zhuang for helpful discussions. The authors are particularly grateful to U. Heinz and C. Shen for developing the VISHNU hydro simulations and for their assistance in using this tool. This material is based upon work supported by the U.S. Department of Energy, Office of Science, Office of Nuclear Physics, within the framework of the Beam Energy Scan Theory (BEST) Topical Collaboration. The work is also supported in part by the National Science Foundation under Grant No. PHY-1352368 (SS and JL), by the National Science Foundation of China under Grant No. 11735007 (JL) and by the U.S. Department of Energy under grant Contract Number No. DE- SC0012704 (BNL)/DE-SC0011090 (MIT) (YY). JL is grateful to  the Institute for Nuclear Theory for hospitality during the INT-16-3 Program.  
The computation of this research was performed on IU's Big Red II cluster, supported in part by Lilly Endowment, Inc. (through its support for the Indiana University Pervasive Technology Institute) and in part by the Indiana METACyt Initiative.


\vspace{-0.2in}

 \begin{thebibliography}{99}



\bibitem{Vilenkin:1980fu} 
  A.~Vilenkin,
  Phys.\ Rev.\ D {\bf 22}, 3080 (1980).
  
\bibitem{Kharzeev:2004ey} 
  D.~Kharzeev,
  Phys.\ Lett.\ B {\bf 633}, 260 (2006). 
  
    \bibitem{Kharzeev:2007tn}
  D.~Kharzeev and A.~Zhitnitsky,
  Nucl.\ Phys.\  A {\bf 797}, 67 (2007).

 \bibitem{Kharzeev:2007jp}
  D.~E.~Kharzeev, L.~D.~McLerran and H.~J.~Warringa,
  Nucl.\ Phys.\  A {\bf 803}, 227 (2008).
  
\bibitem{Fukushima:2008xe} 
  K.~Fukushima, D.~E.~Kharzeev and H.~J.~Warringa,
  Phys.\ Rev.\ D {\bf 78}, 074033 (2008). 
  
   %
\bibitem{Son:2009tf}
  D.~T.~Son and P.~Surowka,
  Phys.\ Rev.\ Lett.\  {\bf 103}, 191601 (2009).

\bibitem{Zakharov:2012vv} 
  V.~I.~Zakharov,
  Lect.\ Notes Phys.\  {\bf 871}, 295 (2013). 
  
\bibitem{Fukushima:2012vr} 
  K.~Fukushima,
  Lect.\ Notes Phys.\  {\bf 871}, 241 (2013). 


\bibitem{Kharzeev:2011ds} 
  D.~E.~Kharzeev and H.~U.~Yee,
  Phys.\ Rev.\ D {\bf 84}, 045025 (2011). 
  
 
 
\bibitem{Nielsen:1983rb} 
  H.~B.~Nielsen and M.~Ninomiya,
  Phys.\ Lett.\  {\bf 130B}, 389 (1983).
  
\bibitem{Son:2012bg} 
  D.~T.~Son and B.~Z.~Spivak,
  Phys.\ Rev.\ B {\bf 88}, 104412 (2013). 


\bibitem{Li:2014bha} 
  Q.~Li {\it et al.},
  Nature Phys.\  {\bf 12}, 550 (2016). 
  
\bibitem{Xiong:2015nna} 
  J.~Xiong {\it et al.},
  arXiv:1503.08179 [cond-mat.str-el].
  
  
  \bibitem{2015PhRvX...5c1023H} Huang, X. {\it et al.}, \ 2015, Physical Review X, 5, 031023.  
  
  \bibitem{2016NatCo...711615A} Arnold, F., Shekhar, C., Wu, S.-C., et al.\ 2016, Nature Communications, 7, 11615.  
  

 
\bibitem{STAR_LPV1} B.~I.~Abelev {\it et al.} (STAR Collaboration), Phys. Rev. Lett. {\bf 103} (2009) 251601.
  
  \bibitem{STAR_LPV2} B.~I.~Abelev {\it et al.} (STAR Collaboration), Phys. Rev. C {\bf 81} (2010) 054908.
  
  \bibitem{STAR_LPV3}
L. Adamczyk {\it et al.} (STAR Collaboration), Phys. Rev. C {\bf 88} (2013) 064911.

\bibitem{STAR_LPV4}
L. Adamczyk {\it et al.} (STAR Collaboration), Phys. Rev. C {\bf 89} (2014) 44908.

\bibitem{STAR_LPV_BES}
L. Adamczyk {\it et al.} (STAR Collaboration), Phys. Rev. Lett. {\bf 113} (2014) 052302.

\bibitem{ALICE_LPV}
B. I. Abelev {\it et al.} (ALICE Collaboration), Phys. Rev. Lett. {\bf 110} (2013) 021301.
  
\bibitem{Kharzeev:2015znc} 
  D.~E.~Kharzeev, J.~Liao, S.~A.~Voloshin and G.~Wang,
  Prog.\ Part.\ Nucl.\ Phys.\  {\bf 88}, 1 (2016). 

\bibitem{Liao:2014ava} 
  J.~Liao,
  Pramana {\bf 84}, no. 5, 901 (2015). 


\bibitem{STAR_LPV_Wang}
Gang Wang (STAR Collaboration),  Nucl. Phys A {\bf 904} (2013) 248c.


\bibitem{Bzdak:2012ia} 
  A.~Bzdak, V.~Koch and J.~Liao,
  Lect.\ Notes Phys.\  {\bf 871}, 503 (2013)
  [arXiv:1207.7327 [nucl-th]].
  


\bibitem{Shen:2014vra} 
  C.~Shen, Z.~Qiu, H.~Song, J.~Bernhard, S.~Bass and U.~Heinz,
  Comput.\ Phys.\ Commun.\  {\bf 199}, 61 (2016). 
 
 
\bibitem{Borsanyi:2011sw} 
  S.~Borsanyi, Z.~Fodor, S.~D.~Katz, S.~Krieg, C.~Ratti and K.~Szabo,
  JHEP {\bf 1201}, 138 (2012). 
  
\bibitem{Bazavov:2012jq} 
  A.~Bazavov {\it et al.},
  Phys.\ Rev.\ D {\bf 86}, 034509 (2012). 
  
  
  \bibitem{Lin:2013sga} 
  S.~Lin and H.~U.~Yee,
  Phys.\ Rev.\ D {\bf 88}, no. 2, 025030 (2013). 
  
 \bibitem{Manuel:2015zpa} 
  C.~Manuel and J.~M.~Torres-Rincon,
  Phys.\ Rev.\ D {\bf 92}, no. 7, 074018 (2015). 
  
  
 
\bibitem{Yin:2015fca} 
  Y.~Yin and J.~Liao,
  Phys.\ Lett.\ B {\bf 756}, 42 (2016). 

\bibitem{Yee:2013cya} 
  H.~U.~Yee and Y.~Yin,
  Phys.\ Rev.\ C {\bf 89}, no. 4, 044909 (2014). 
  
 

\bibitem{Bloczynski:2012en} 
  J.~Bloczynski, X.~G.~Huang, X.~Zhang and J.~Liao,
  Phys.\ Lett.\ B {\bf 718}, 1529 (2013). 

\bibitem{McLerran:2013hla} 
  L.~McLerran and V.~Skokov,
  Nucl.\ Phys.\ A {\bf 929}, 184 (2014). 
  
\bibitem{Gursoy:2014aka} 
  U.~Gursoy, D.~Kharzeev and K.~Rajagopal,
  Phys.\ Rev.\ C {\bf 89}, no. 5, 054905 (2014). 
  
  
\bibitem{Tuchin:2015oka} 
  K.~Tuchin,
  Phys.\ Rev.\ C {\bf 93}, no. 1, 014905 (2016). 
  
\bibitem{Kharzeev:2001ev} 
  D.~Kharzeev, A.~Krasnitz and R.~Venugopalan,
  Phys.\ Lett.\ B {\bf 545}, 298 (2002). 
  
\bibitem{Muller:2010jd} 
  B.~Muller and A.~Schafer,
  Phys.\ Rev.\ C {\bf 82}, 057902 (2010). 
     
     
\bibitem{Hirono:2014oda} 
  Y.~Hirono, T.~Hirano and D.~E.~Kharzeev,
  arXiv:1412.0311 [hep-ph].
  
\bibitem{Mace:2016svc} 
  M.~Mace, S.~Schlichting and R.~Venugopalan,
  Phys.\ Rev.\ D {\bf 93}, no. 7, 074036 (2016). 
 
\bibitem{Rezaeian:2012ji} 
  A.~H.~Rezaeian, M.~Siddikov, M.~Van de Klundert and R.~Venugopalan,
  Phys.\ Rev.\ D {\bf 87}, no. 3, 034002 (2013). 
  
\bibitem{Schenke:2012wb} 
  B.~Schenke, P.~Tribedy and R.~Venugopalan,
  Phys.\ Rev.\ Lett.\  {\bf 108}, 252301 (2012). 
  
\bibitem{Kowalski:2007rw} 
  H.~Kowalski, T.~Lappi and R.~Venugopalan,
  Phys.\ Rev.\ Lett.\  {\bf 100}, 022303 (2008). 
  
  
\bibitem{Hongo:2013cqa} 
  M.~Hongo, Y.~Hirono and T.~Hirano,
  arXiv:1309.2823 [nucl-th].
  
\bibitem{Denicol:2012vq} 
  G.~S.~Denicol, H.~Niemi, I.~Bouras, E.~Molnar, Z.~Xu, D.~H.~Rischke and C.~Greiner,
  Phys.\ Rev.\ D {\bf 89}, no. 7, 074005 (2014). 
 
 
\bibitem{Voloshin:2004vk} 
  S.~A.~Voloshin,
  Phys.\ Rev.\ C {\bf 70}, 057901 (2004). 
  

 
\bibitem{Bloczynski:2013mca} 
  J.~Bloczynski, X.~G.~Huang, X.~Zhang and J.~Liao,
  Nucl.\ Phys.\ A {\bf 939}, 85 (2015). 
  
\bibitem{Skokov:2016yrj} 
  V.~Skokov, P.~Sorensen, V.~Koch, S.~Schlichting, J.~Thomas, S.~Voloshin, G.~Wang and H.~U.~Yee,
  arXiv:1608.00982 [nucl-th].
  
\bibitem{Deng:2016knn} 
  W.~T.~Deng, X.~G.~Huang, G.~L.~Ma and G.~Wang,
  Phys.\ Rev.\ C {\bf 94}, 041901 (2016). 
  
  
\bibitem{Wang:2016iov} 
  F.~Wang and J.~Zhao,
  arXiv:1608.06610 [nucl-th].
  
\bibitem{Wen:2016zic} 
  F.~Wen, L.~Wen and G.~Wang,
  arXiv:1608.03205 [nucl-th].


\bibitem{Ma:2011uma} 
  G.~L.~Ma and B.~Zhang,
  Phys.\ Lett.\ B {\bf 700}, 39 (2011).
  
  
\bibitem{Mueller:2016ven} 
  N.~Müller, S.~Schlichting and S.~Sharma,
  Phys.\ Rev.\ Lett.\  {\bf 117}, no. 14, 142301 (2016). 
  
     
  \bibitem{Burnier:2011bf}
  Y.~Burnier, D.~E.~Kharzeev, J.~Liao and H.~U.~Yee,
  Phys.\ Rev.\ Lett.\  {\bf 107}, 052303 (2011); arXiv:1208.2537 [hep-ph].
  

\bibitem{Adamczyk:2015eqo} 
  L.~Adamczyk {\it et al.} [STAR Collaboration],
  Phys.\ Rev.\ Lett.\  {\bf 114}, no. 25, 252302 (2015). 
  
   
\end{thebibliography}
\end{document}